\documentclass[aps,prl,floats,twocolumn,amsfonts,amssymb,unsortedaddress,floatfix]{revtex4}
\usepackage{epsfig}
\usepackage{amssymb}
\usepackage{mathtools}
\usepackage{bm}
\usepackage{color}

\def\be{\begin{equation}}       \def\ee{\end{equation}}
\def\bea{\begin{eqnarray}}      \def\eea{\end{eqnarray}}

\begin{document}
\title{Crossover from Jamming to Clogging Behaviors in Heterogeneous Environments} 

\author{Huba P\'eter$^{1,2}$, Andr\'as Lib\'al$^{1,2}$,  Charles Reichhardt$^1$, and Cynthia J. O. Reichhardt$^{1,*}$, 
 }

\affiliation{$^1$Theoretical Division, Los Alamos National Laboratory, Los Alamos, NM, 87545, USA}  
\affiliation{$^{2}$Mathematics and Computer Science Department, Babe{\c s}-Bolyai University, Cluj, 400084, Romania }

\begin{abstract}
\end{abstract}

\maketitle

{\bf
Jamming describes a transition from a flowing or liquid state to a solid or rigid state in a  loose assembly of particles such as grains or bubbles.  In contrast, clogging describes the ceasing of the flow of particulate matter through a bottleneck.  It is not clear how to distinguish jamming from clogging, nor is it known whether they are distinct phenomena or fundamentally the same.  We examine an assembly of disks moving through a random obstacle array and identify a transition from clogging to jamming behavior as the disk density increases. The clogging transition has characteristics of an absorbing phase transition, with the disks evolving into a heterogeneous phase-separated clogged state after a critical diverging transient time.  In contrast, jamming is a rapid process in which the disks form a homogeneous motionless packing, with a rigidity length scale that diverges as the jamming density is approached.
}

The concept of jamming
is used
in loose assemblies of particles such as grains or bubbles
to describe
the transition
from an easily flowing
fluidlike state
to a rigid jammed or solidlike state 
\cite{1,2,3,4}.
Liu and Nagel proposed a generalized 
jamming phase diagram combining 
temperature, load, and density, where a particularly important point is 
the density $\phi_j$ at
which jamming occurs \cite{1}.
Jamming has been extensively  
studied in a variety of systems \cite{3,4,5}, and there is evidence
that in certain cases, 
the jamming transition has the properties of 
a critical point, such as a correlation length that diverges as the jamming 
density is approached \cite{2,3,4,5,16,17,18,19}. 
A related phenomenon is the clogging that occurs for particles
flowing through a hopper, where as a function of time there is a probability for
arch structures to form that block the flow \cite{6,7,8,9}.
Clogging is associated with the motion of particulate matter past
a physical constraint such as wells, barriers, obstacles, 
or bottlenecks \cite{10,11,12,13,14,15};
however, it has not been established whether jamming and clogging
are two forms of the same phenomenon
or whether there are key features that distinguish jamming from clogging.

Here we show for frictionless disks moving
through a random obstacle array that jamming and clogging
are distinct phenomena and that a transition from
clogging to jamming
occurs as a function of increasing disk density.
We
identify the number of 
obstacles required to stop the flow and
the transient times needed to reach a stationary clogged or jammed state.
In the jamming regime,
the
obstacle density $\phi_c^j$ at which flow ceases
deceases with increasing 
disk density,
and the system forms a homogeneous jammed state when the rigidity correlation
length associated with
$\phi_j$
becomes larger than the average distance between obstacles. 
In contrast, during clogging
the system organizes over time
into a heterogeneous or phase-separated state,
and the transient time diverges at a critical obstacle density
$\phi_c^c$ that is independent
of the disk density.
The phase-separated state consists of
regions with a density near $\phi_j$ coexisting with
low density regions.   

 \begin{figure*}
\begin{center}
\includegraphics[width=.7\textwidth]{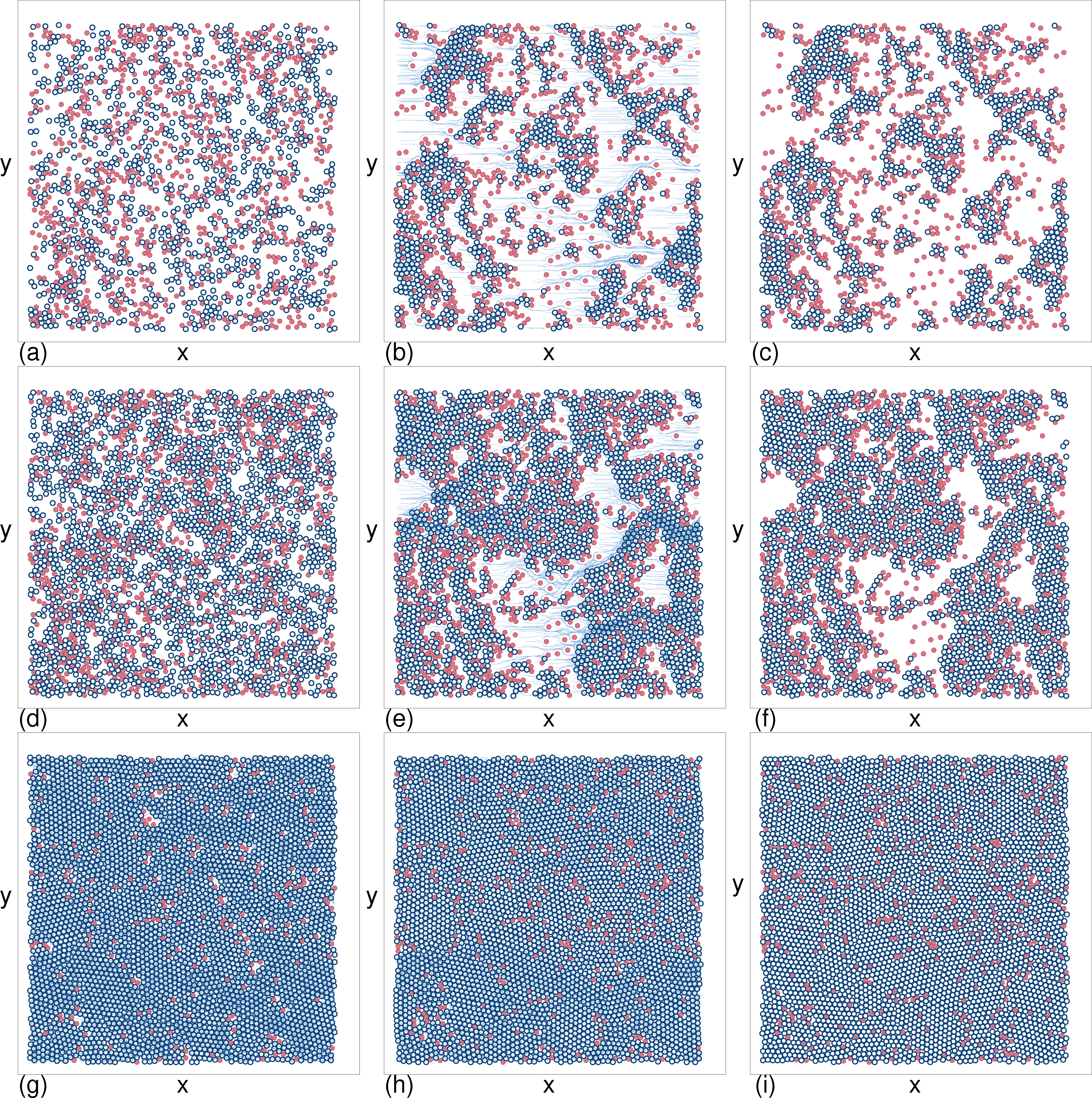}
\caption{{\bf Clogging and jamming in obstacle arrays.}
  Images of ({\bf a}) initial, ({\bf b}) transient flowing, and ({\bf c}) final clogged state for
  mobile disks (dark blue open circles) driven
  in the positive $x$ direction through obstacles (red filled circles) in
  a sample with disk density $\phi_p=0.2186$ and obstacle density $\phi_{ps}=0.175$.
  Light blue lines indicate the disk trajectories over a fixed time period.
  The disks are initially in a flowing state and evolve into a phase-separated clogged state.
  ({\bf d}) Initial, ({\bf e}) transient flowing, and ({\bf d}) final clogged state images for
  a sample with higher disk density $\phi_{p} = 0.436$
  and $\phi_{ps} = 0.175$,
  which also begins in a uniform flowing state and evolves toward a
  phase-separated clogged state.  The dense regions have a local disk density of
  $\phi_l=0.84$.
  ({\bf g}, {\bf h}, {\bf i}) Jamming under increasing obstacle density in a sample
  with $\phi_p=0.785$.
  ({\bf g}) At an obstacle density of $\phi_{ps} = 0.043$,
  we find steady state flow.
  ({\bf h}) At $\phi_{ps}=0.065$ the steady state flow is reduced but still
  present.
  ({\bf i}) At $\phi_{ps} = 0.0872$, the system jams.
  The jammed state is much more uniform in density than the clogged state, and
  jamming occurs rapidly with almost no transient flow
  above a critical obstacle density $\phi_{c}^j$.
}
\label{fig:1}
\end{center}
\end{figure*}

\section{Results}
{\bf Time evolution to a jammed or clogged state.}
We numerically examine
disks driven through a two-dimensional array of obstacles in the form of
immobile disks.
The total area density of the system is $\phi_{\rm tot}=\phi_p + \phi_{ps}$, where
$\phi_p$ is the area density of the moving disks and $\phi_{ps}$ is the area density
of the obstacles.
Starting from a uniformly dense sample, we apply a driving force and find that
over time the system evolves either to
a steady free flowing state or to
a motionless clogged or jammed state.
In Fig.~\ref{fig:1}a,b,c we illustrate
the time evolution of a system
with a disk density of $\phi_{p} =  0.2186$ and
an obstacle density of $\phi_{ps} = 0.175$,
beginning with the uniform density initial state in Fig.~\ref{fig:1}a.
Upon application of a drive, we find a transient flowing state
as shown in Fig.~\ref{fig:1}b which gradually evolves into the final motionless
phase-separated or clustered clogged state in Fig.~\ref{fig:1}c.
For a higher disk density of $\phi_p=0.436$,
Fig.~\ref{fig:1}d,e,f shows that the same evolution from uniform initial state
to transient flowing state to static clogged state occurs, but the dense clusters in
the clogged state are larger.
At much higher disk densities of $\phi_p=0.785$, we find jamming behavior
when the obstacle density is larger than a critical value $\phi_c^j$.
Below $\phi_c^j$, the system quickly settles into steady state flow, as
shown in Fig.~\ref{fig:1}g for $\phi_{ps}=0.043$ and in Fig.~\ref{fig:1}h
for $\phi_{ps}=0.065$.  The magnitude of the flow decreases with increasing
$\phi_{ps}$.  Above $\phi_c^j$ the disks quickly form a jammed state when
driven, as illustrated in Fig.~\ref{fig:1}i for $\phi_{ps}=0.0872$.  In contrast to
the
density phase-separated clogged states that form at lower $\phi_p$, the jammed states are
homogeneously dense.

\begin{figure}
\begin{center}
\includegraphics[width=1.0\columnwidth]{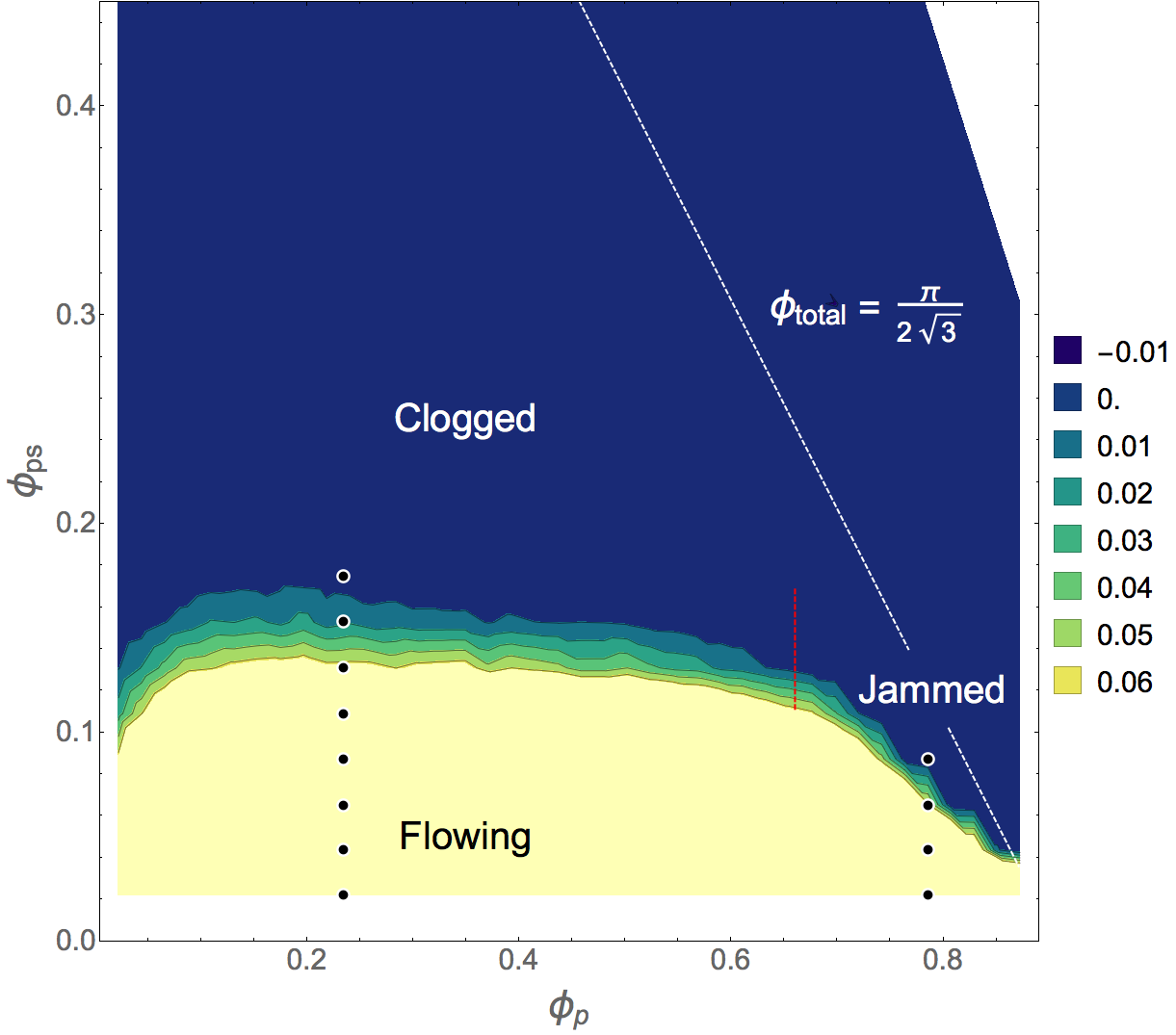}
\caption{{\bf Clogging-jamming phase diagram.} 
  The heat map of the disk velocity $V_0$ after $10^6$
  simulation time steps as a function of
  obstacle density $\phi_{ps}$ vs disk density $\phi_p$.
  Yellow indicates high $V_0$ and blue indicates zero $V_0$.
  The white dashed line is the density
  $\phi_{\rm tot} = \pi/2\sqrt{3} \approx 0.9069$ at which
  the disks would form a hexagonal solid. 
Clogging occurs for
$\phi_{p} < 0.67$, and the critical
obstacle density for clogging is nearly independent of $\phi_p$,
$\phi_{c}^c \approx 0.153$.
Jamming occurs for $\phi_{p} > 0.67$, as indicated by the red vertical dashed line,
and $\phi_c^j$, the critical obstacle density
for jamming, decreases linearly with increasing $\phi_{p}$.
The dots along $\phi_{p}=0.234$ indicate the values of $\phi_{ps}$ shown in the time
series of Fig.~\ref{fig:3}a, while the dots along
$\phi_{p}=0.785$ indicate the values of $\phi_{ps}$ shown in the time
series of Fig.~\ref{fig:3}d.
Above the dashed white line is a region in which no data can be taken.
}
\label{fig:2}
\end{center}
\end{figure}

{\bf Velocity measurement of the transition from clogging to jamming.}
To characterize the system we perform a series of simulations with varied $\phi_{p}$ and 
$\phi_{ps}$.
We measure the final velocity $V_0$ of the mobile disks after
a fixed time interval, and average over ten disorder realizations.
In Fig.~\ref{fig:2} we plot a
velocity heat map
as a function of $\phi_{ps}$ versus $\phi_{p}$.
The dashed white line
indicates the density $\phi_{ps}$ above which
$\phi_{\rm tot} =\phi_j= \pi/2\sqrt 3 \approx 0.9069$, where
a dense or jammed hexagonal disk packing would form.
We find a flowing regime at small $\phi_{ps}$, a clogged regime
for $\phi_{p} < 0.67$, and a jammed regime for
$\phi_{p} > 0.67$. 
The critical obstacle density
$\phi_c^c$ above which the velocity $V_0$ drops to zero in the clogging regime
remains roughly constant at $\phi_c^c\approx 0.153$, independent of the
value of $\phi_p$.
In the jamming regime, the critical obstacle density $\phi_c^j$ separating flowing from
jammed states
decreases linearly with increasing $\phi_{p}$
and reaches $\phi_c^j=0$ for $\phi_{p} \approx 0.9069$.   
This indicates that the transition
to a clogged state is controlled by the average spacing
$l_{ps}=1/\sqrt{\phi_{ps}}$ between
obstacles, similar to the manner in which hopper clogging is controlled by the aperture
size.
In contrast, the transition to a jammed state
is controlled by a growing correlation length $\xi$ associated with
the jamming or crystallization point $\phi_j$ \cite{5}.
We argue that the system jams when $\xi=l_{ps}$.
If we assume that near jamming in a clean system, 
the correlation length grows as
$\xi \propto (\phi_{j} - \phi_p)^{-\nu}$,
then the transition to the jammed state varies with obstacle density
according to
$\phi_c^j \propto (\phi_{j} -\phi_{p})^{2\nu}$.
In Fig.~\ref{fig:2}, $\phi_c^j \propto \phi_p$, implying
that $\nu=1/2$, consistent with the exponent
$\nu = 1/2$
proposed for jamming in Refs.~\cite{20,21},
as well as with simulation measurements giving
$\nu$ in the range 0.6 to 0.7
for two-dimensional bidisperse disks
\cite{16,18}. The exponent we find 
is also in agreement with
that observed for the shift in the jamming  point in bidisperse disks
on random pinning arrays \cite{22}.
Studies of
bidisperse disk jamming with dilute
obstacles very near $\phi_j$
also show that
$\phi_j$ decreases linearly with obstacle density,
giving $\nu = 1/2$ \cite{L1}.  

Previous simulations of bidisperse disks
of radius $R_s=0.5$ and $R_l=0.7$ flowing through a periodic array of
obstacles with radius $R_s=0.5$
 showed that
clogging is strongly enhanced
when $l_{ps} \lesssim 2.35$ \cite{23}.
This is because in order for a pair of disks, one large and one small,
to fit between two obstacles,
the lattice constant $a$ of the obstacle array must be at least
$4R_s+2R_l=2.4$.
In our monodisperse disk system, the obstacles are placed randomly, but one can
obtain an estimate of the $l_{ps}$ for the onset of clogging by considering the
holes in the obstacle array \cite{1986hinrichsen}.
The minimum number $n$ of holes
of size $R$ per obstacle required to prevent the obstacle arrangement from becoming
anisotropic is $n=5$.
For a pair of disks to pass between two obstacles, the obstacle spacing
must be at least $l_{ps}=6R=3.0$.  This spacing can be achieved by
placing the obstacles such that the holes cannot overlap,
giving an effective obstacle radius of $3R$ and an obstacle
density of $\phi_{ps}=(1/6)(\pi/(2\sqrt{3}))=0.15$.
In Fig.~\ref{fig:2}, the onset of clogging,
$\phi_{c}^c\approx 0.153$,
is close to this density.
When $\phi_p\lesssim 0.15$,
$\phi_c^c$ is no longer constant but decreases with decreasing $\phi_p$.
At these low disk densities, mobile disks are trapped independently,
so at least one additional obstacle must be added for every additional
mobile disk, giving $\phi_c^c \propto \phi_p$.

\begin{figure*}
\begin{center}
\includegraphics[width=1.4\columnwidth]{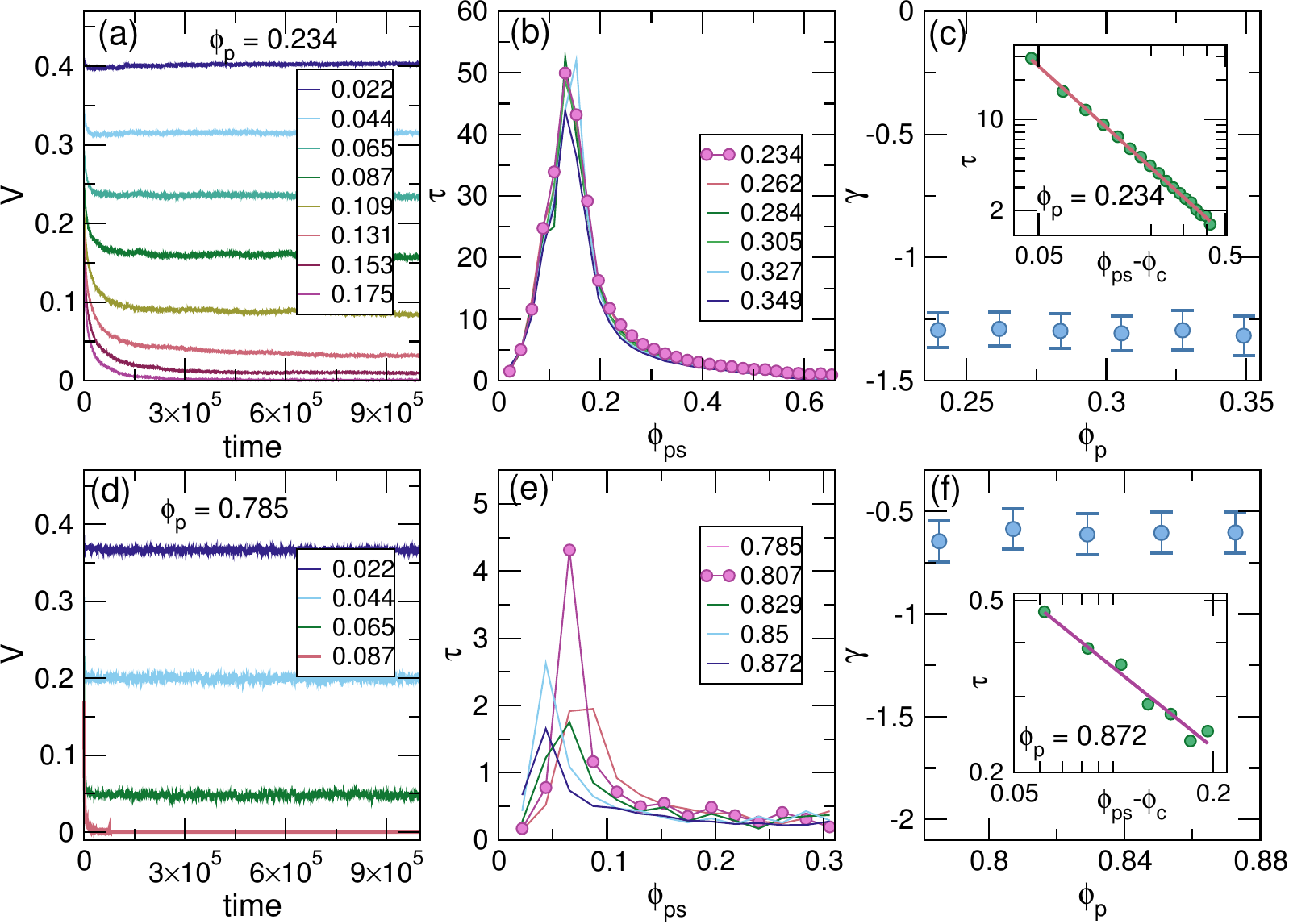}
\caption{{\bf Transient times for clogging and jamming.}
  ({\bf a}) The average velocity $V$ per mobile disk vs time
  in simulation time steps for
  samples with mobile disk density $\phi_{p} = 0.234$ at varied
  obstacle density $\phi_{ps}=0.022$ to 0.175, from top to
  bottom.  The transient time decreases with increasing $\phi_{ps}$, and
  the disks reach a clogged state for $\phi_{ps}>0.153$.
  ({\bf b}) Transient times $\tau$ vs $\phi_{ps}$ obtained from
  the curves in {\bf a}
  by fitting $V \propto A\exp(-t/\tau) +V_0$
  for $\phi_p=0.234$ to 0.349, from top to bottom.
  There is a divergence in $\tau$
  near the clogging density of 
  $\phi_{c}^c = 0.153$.
  ({\bf c})
  The value of the exponent $\gamma$ vs $\phi_p$ obtained by fitting
  the curves in {\bf b} to
  $\tau \propto (\phi_{ps} - \phi_{c}^c)^{\gamma}$.
  Inset: $\tau$ vs $\phi_{ps}-\phi_c^c$ at $\phi_p=0.234$.
  The pink line indicates a power law fit
  with $\gamma = -1.29 \pm 0.1$.
  ({\bf d}) $V$ vs time for samples with $\phi_p=0.785$ in the jamming regime
  for varied $\phi_{ps}=0.022$ to 0.087 from top to bottom.  
  ({\bf e}) $\tau$ vs $\phi_{ps}$ obtained from the curves in {\bf d} for
  $\phi_p=0.785$ to 0.872, from top to bottom.
  The transient times are much shorter than those in the clogging regime
  in panel {\bf b}.
  ({\bf f}) Exponent $\gamma$ vs $\phi_p$ obtained by
  fitting the curves in {\bf e} to
  $\tau \propto (\phi_{ps} - \phi_{c}^j)^{\gamma}$.
  Inset: $\tau$ vs $\phi_{ps}-\phi_c^j$ at $\phi_p=0.872$.
  The pink line indicates a power law fit with 
  $\gamma = -0.66$. 
}
\label{fig:3}
\end{center}
\end{figure*}

{\bf Transient velocities near the clogging and jamming transitions.}
In Fig.~\ref{fig:3}a we show representative time series of the
average velocity $V$ per mobile disk in the clogging regime
at $\phi_{p} = 0.234$ for $\phi_{ps}$ ranging from $\phi_{ps}=0.022$ to 0.175.
At low obstacle densities
such as $\phi_{ps} = 0.022$ and $\phi_{ps}=0.044$,
the disks reach a steady state flow after a very short transient time $\tau$.
As $\phi_{ps}$ increases, $\tau$ increases, showing a divergence
at the critical obstacle density $\phi_c^c$ where clogging first occurs,
while for $\phi_{ps}>\phi_c^c$, $\tau$ decreases
with increasing $\phi_{ps}$ and the
disks reach a completely clogged state with $V=0$.
We fit
$V(t) \propto A\exp(-t/\tau) + V_{0}$, and plot the resulting values of
$\tau$
in Fig.~\ref{fig:3}b as a function of $\phi_{ps}$ for $\phi_p=0.234$ to 0.349.
In each case, $\tau$ diverges near
$\phi_{ps} = 0.15$.
We fit this divergence
for $\phi_{ps}>\phi_c^c$ to a power law,
$ \tau \propto (\phi_{ps} - \phi_{c}^c)^\gamma$,
as shown in the inset of Fig.~\ref{fig:3}c
for $\phi_{p} = 0.234$, 
where $\gamma = -1.29 \pm 0.1$.
The plot of $\gamma$ versus $\phi_p$ in the main
panel of Fig.~\ref{fig:3}(c) indicates that
$\gamma$ has a constant value in the range $-1.25$ to $-1.35$.
The transient time behavior and exponent values are similar to those
found for the
diverging time scales that appear
near the
irreversible-reversible transition in systems
exhibiting random organization \cite{24,N,N2} and near the depinning transition
for colloids \cite{25} and vortices \cite{26,27} driven over random pinning arrays.  
The power law exponents are also close to the value $\gamma=-1.295$
expected for the universality class of two-dimensional directed percolation 
\cite{28},
and we find similar values of $\gamma$ for $\phi_p<0.67$
throughout the clogging regime.
Directed percolation is often used to 
describe nonequilibrium absorbing phase transitions \cite{28}, 
and in our case the steady state flow corresponds
to a fluctuating state,
while the clogged state is the non-fluctuating or absorbed state.   

In the jamming regime the transient times are much shorter, as shown by
the plot of $V(t)$ in Fig.~\ref{fig:3}d
for $\phi_p=0.785$.
In Fig.~\ref{fig:3}e,
$\tau$ versus $\phi_{ps}$ in the range $\phi_p=0.785$ to 0.872 has
a value that is
an average of 20 times smaller than in the clogging regime from
Fig.~\ref{fig:3}b.
The peak in $\tau$ shifts to lower $\phi_{ps}$ with increasing $\phi_p$, reflecting
the behavior of the critical jamming density $\phi_c^j$.
By fitting the curves in Fig.~\ref{fig:3}e to
$\tau \propto (\phi_{ps}-\phi_c^j)^\gamma$,
as demonstrated in the inset of Fig.~\ref{fig:3}f for $\phi_p=0.872$,
we obtain
$\gamma\approx -0.66$, as shown in the plot of $\gamma$ versus $\phi_p$ in the
main panel of Fig.~\ref{fig:3}f.
This indicates that there is 
a pronounced change in the dynamics of the jamming regime
compared to the clogging regime.

 \begin{figure}
\begin{center}
\includegraphics[width=1.0\columnwidth]{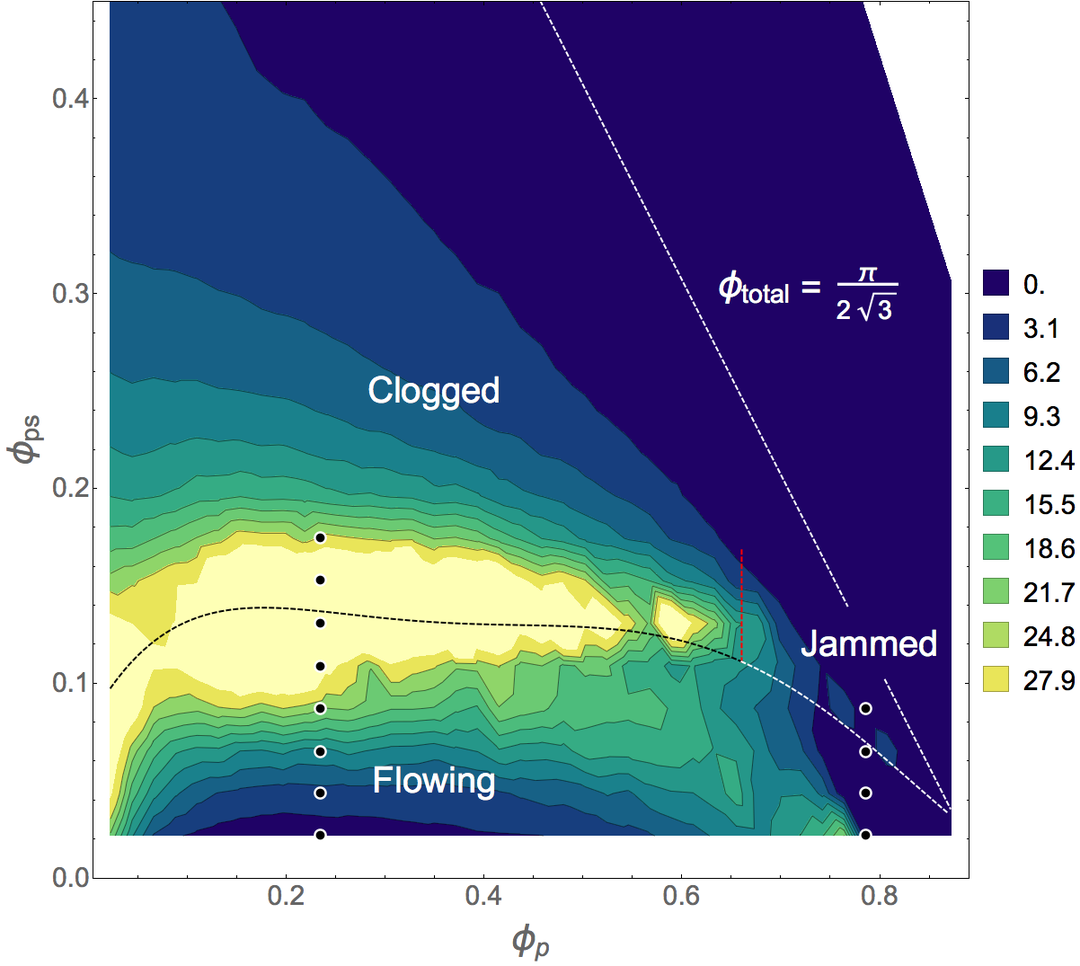}
\caption{{\bf Transient time behavior.}
  The heat map of the transient times $\tau$
  obtained from fitting $V(t)=A\exp(-t/\tau) + V_{0}$ as a function of
  $\phi_{ps}$ vs $\phi_{p}$.  Yellow indicates large $\tau$ and blue indicates
  small $\tau$.
  The dark dashed line is a guide to the eye marking the crossover from
  a flowing  state to a clogged state,
  while the lower dashed white line indicates the transition from a flowing
  state to a jammed state.  The upper dashed line is the crystallization
  density $\phi_{\rm tot}=\pi/2\sqrt{3}$, and no data was obtained
  above this line.
  The dots along $\phi_{p} = 0.234$ indicate the values of $\phi_{ps}$ shown
  in the time series of Fig.~\ref{fig:3}a,
  while the dots along $\phi_p=0.785$ indicate the values of $\phi_{ps}$ shown
  in the time series of Fig.~\ref{fig:3}d.
  The system must organize into a clogged state, giving large transient times in the
  clogging regime, but can quickly enter a jammed state, giving small transient
  times in the jamming regime.
}
\label{fig:4}
\end{center}
\end{figure}

 In Fig.~\ref{fig:4} we show a heat map of the transient time
 $\tau$ obtained by fitting 
 $V(t)=A\exp(-t/\tau) + V_{0}$. 
The transient times become large
near the crossover from flowing to clogging
for $\phi_{p} < 0.67$, while in the jamming regime for $\phi_{p} > 0.67$,
the transient times are strongly reduced.
This provides further evidence that in the clogging regime it is necessary
for the system to organize over time into a clogged state,
gradually forming phase-separated regions of high and low density as
illustrated in Fig.~\ref{fig:1}a-f.
In contrast, the jammed system has strong spatial correlations,
and once the correlation length
associated with $\phi_j$
is larger than the distance $l_{ps}$ between defects,
very few disk rearrangements are needed to bring the system
into a stationary, nonflowing state.

\begin{figure}
\begin{center}
\includegraphics[width=0.9\columnwidth]{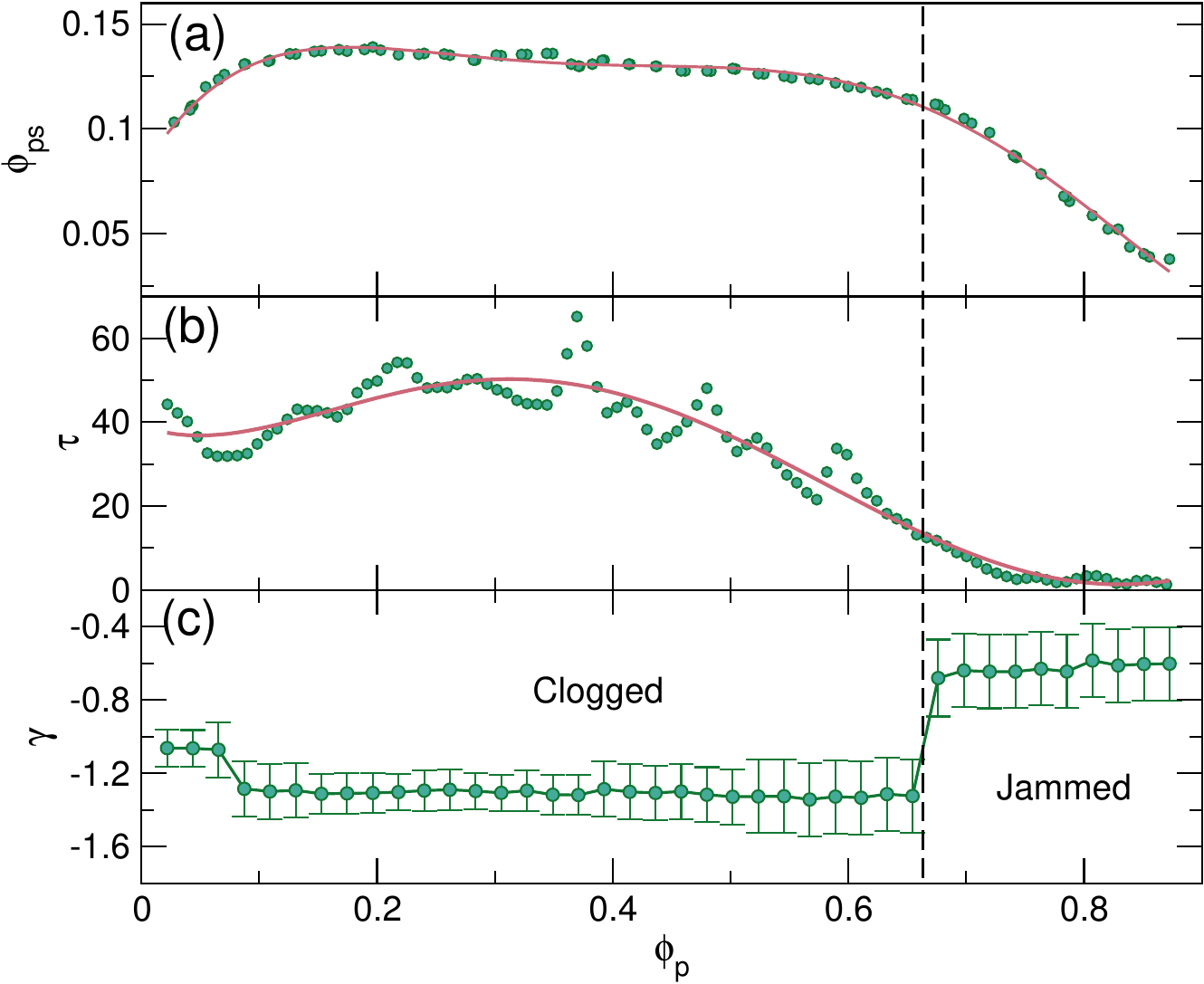}
\caption{{\bf Transient times and critical exponents across the clogging to jamming
transition.}
  ({\bf a}) The location of the transition from a flowing state to a clogged
  or jammed state, 
  defined as points for
  which $V_0=0.01$,
  as a function of $\phi_{ps}$ vs $\phi_p$.
  The dashed line separates clogged states at low $\phi_p$ from
  jammed states at high $\phi_p$.
  ({\bf b}) The transient time $\tau$ at the flowing to nonflowing transition point
  vs $\phi_p$.
  ({\bf c}) The transient exponent $\gamma$ extracted from the nonflowing side
  of the transition vs $\phi_p$.
  There is a clear crossover from clogging to jamming.
  In the clogging regime, $\gamma \approx -1.29$, but in the jamming regime,
  $\gamma \approx -0.66$, indicating that the dynamics of clogging differ from those
  of jamming.}
\label{fig:5}
\end{center}
\end{figure}

In Fig.~\ref{fig:5}a we show the transition from the flowing to the clogged or jammed
state as a function of $\phi_{ps}$ versus $\phi_p$ by identifying the points from
Fig.~\ref{fig:2} for which $V=0.01$.
Figure~\ref{fig:5}b shows the transient times $\tau$ along this transition line,
and in Fig.~\ref{fig:5}c we plot
the transient exponent $\gamma$.
The dashed vertical line at $\phi_p=0.67$ indicates a transition from clogging
to jamming behavior, correlated with a change from
$\gamma \approx -1.29$ in the clogging regime to $\gamma \approx -0.66$ in the
jamming regime, as well as with a drop in $\phi_{ps}$ and $\tau$.
The point $\phi_{p} = 0.67$
can be interpreted as the density at which
the correlation length associated with
the jamming or crystallization density
drops below $4R_d$, corresponding to two disk diameters.
We find a third value of $\gamma$ for $\phi_{p} < 0.07$
in a density regime where
the value of $\phi_{ps}$ at which a clogged state appears decreases with
decreasing $\phi_p$.
This regime is dominated by
the trapping of single disks rather than collective clogging
dynamics.

\begin{figure*}
\begin{center}
\includegraphics[width=1.2\columnwidth]{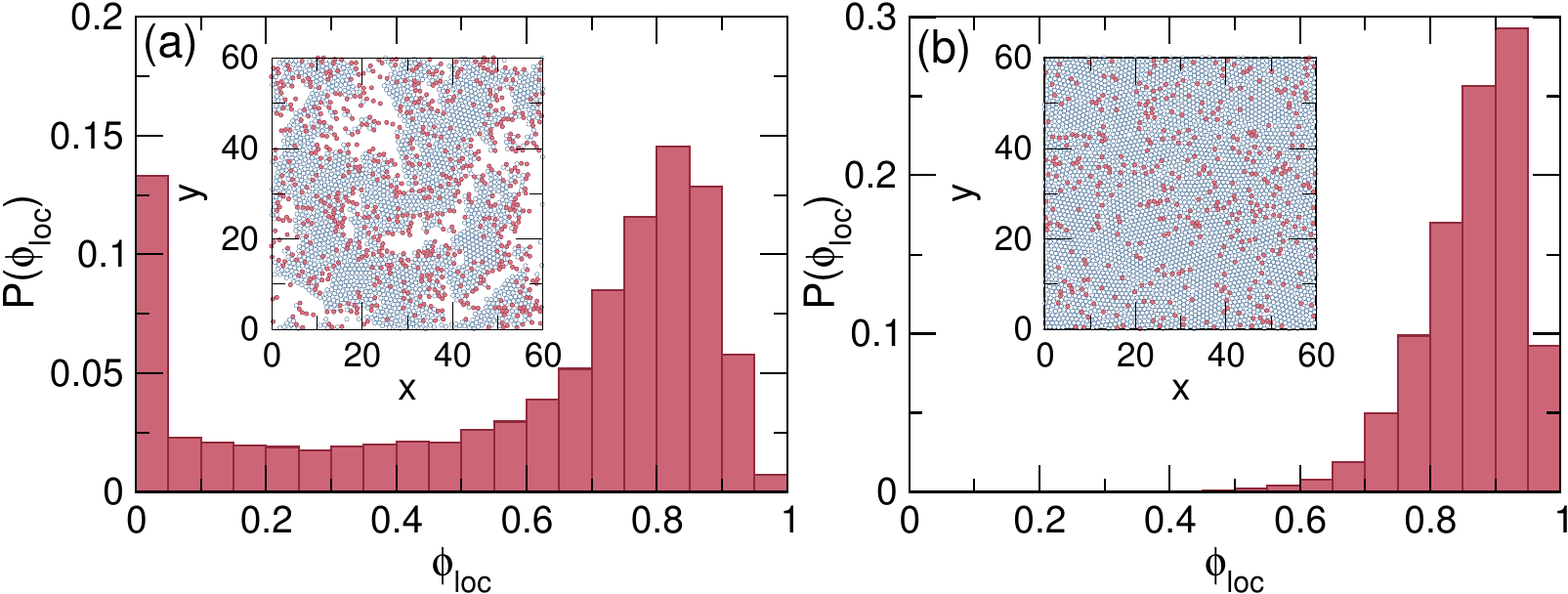}
\caption{{\bf Local density distributions in clogged and jammed systems.}   
  ({\bf a}) The local density distribution $P(\phi_{\rm loc})$
  averaged over 10 realizations
  for a system
  with $\phi_{p} = 0.5$ and $\phi_{ps} = 0.175$
  that reaches a clogged state in which the local density is bimodally distributed.
  Inset:  Image of a clogged configuration from one of the realizations,
  showing the mobile disks (dark blue open circles) trapped by the obstacles
  (red filled circles).
  (b) $P(\phi_{\rm loc})$ averaged over ten realizations
  for a jammed system with $\phi_{p} = 0.8$ and
  $\phi_{ps} = 0.06$ shows a single peak at $\phi_{\rm tot}$.
  Inset: Image of a jammed configuration from one of the realizations.
}
\label{fig:6}
\end{center}
\end{figure*}
 
{\bf Local disk densities in clogged and jammed states.}
The clogged and jammed systems can also be distinguished
by examining the local disk density $\phi_{\rm loc}$ measured in
areas $6R_d \times 6R_d$ in size.
In Fig.~\ref{fig:6}a we plot the local 
density distribution $P(\phi_{\rm loc})$
averaged over ten realizations
of the final clogged state
for a system with $\phi_{p} = 0.5$ and $\phi_{ps} = 0.175$.
As shown in the inset of Fig.~\ref{fig:6}a,
the disks phase separate into low density regions associated with the
peak at $\phi_{\rm loc}=0.1$ and high density regions which produce a
second peak at $\phi_{\rm loc}=0.85$.
The local density of the dense regions is lower than the value of
$\phi_{\rm loc}=0.9069$ for a dense ordered hexagonal disk arrangement due to
the considerable disorder introduced in the packing by the randomly placed obstacles.
In Fig.~\ref{fig:6}b, $P(\phi_{\rm loc})$ for
a system with
$\phi_{p} = 0.8$ and $\phi_{ps} = 0.06$ that reaches a jammed state
has a single peak near $\phi_{\rm loc}=0.9$, reflecting the uniform
disk density at jamming that is illustrated in the inset of Fig.~\ref{fig:6}(b).

\section{Discussion}
Our results suggest that clogging and jamming processes have different dynamics.
Clogging in the presence of random obstacles has signatures of an
absorbing transition falling in a directed percolation universality
class, and its dynamics are controlled by the average spacing of the obstacles.
In the jamming that occurs for higher $\phi_{\rm tot}$,
the dynamics are controlled by the growing correlation length associated
with $\phi_j$, the jamming or crystallization density of an obstacle-free system.
These results show that jamming and clogging
in obstacles are indeed different phenomena.
Jamming
is associated with an equilibrium critical point,
the formation of a homogeneous rigid state, and short transient
times to reach this state,
while clogging is a nonequilibrium dynamical phenomenon in which
the system evolves over an extended time into a
strongly spatially heterogeneous state.
Our results have implications 
for flow though heterogeneous media \cite{29}, erosion \cite{30},
depinning transitions in particle assemblies \cite{31},
and active matter in disordered environments \cite{32,33}.
Experimentally our results could be tested using colloidal particles
at low flow rates to reduce hydrodynamic effects. 
It would also be interesting to examine the effects of adding
frictional contacts between the disks, since these
can change the characteristics of the jamming transition \cite{34,35},
or to replace the disks by elongated particles \cite{36} or chains \cite{37,38}.

This work was carried out under the auspices of the National Nuclear
Security Administration of the U.S. Department of Energy at Los Alamos
National Laboratory under Contract No. DEAC52-006NA25396.  The authors
wish to thank LDRD at LANL for financial support through grant
20170147ER.

{\bf Author contributions:} C.R. and C.J.O.R. conceived the work and wrote
the manuscript, H.P. and  A.L. conducted the simulations and performed data
analysis.

\section{Methods}
{\bf Numerical simulation.}
We conduct simulations of
nonoverlapping disks and obstacles confined to
a two-dimensional plane. The system size is $L \times L$ with $L=60$, and
we use periodic boundary conditions in both the $x$ and $y$ directions. 
We introduce $N_p$ mobile disks of radius $R_d=0.5$ along with $N_{ps}$ obstacles
represented by disks of radius $R_d$
that are not allowed to move.
The area coverage of the mobile disks is $\phi_p=N_p\pi R_d^2/L^2$, the
area coverage of the obstacles is $\phi_{ps}=N_{ps}\pi R_d^2/L^2$, and
the total area coverage is $\phi_{\rm tot}=\phi_p+\phi_{ps}$.

The
disk dynamics are given by the overdamped equation of motion
\begin{equation}
\frac{1}{\eta}\frac{\Delta {\bf r}_i}{\Delta t} =  {\bf F}_{pp}^i + {\bf F}_{\rm obs}^i + {\bf F}_{d}
\end{equation}
where $\eta=1$ is the viscosity.
The interaction between two disks at ${\bf r}_i$ and ${\bf r}_j$ is
a short range harmonic repulsion,
${\bf F}_{dd}^{ij}=k(r_{ij}-2R_d)\Theta(r_{ij}-2R_d){\bf {\hat r}}_{ij}$,
where $r_{ij}=|{\bf r}_i-{\bf r}_j|$, ${\bf {\hat r}}_{ij}=({\bf r}_i-{\bf r}_j)/r_{ij}$, and
$\Theta$ is the Heaviside step function.
We set $k = 200$, which is large enough that overlap
between disks does not exceed $0.01R_d$, placing us  in the hard disk limit.
The interactions with mobile disks are given by
${\bf F}_{pp}^i=\sum_{j\neq i}^{N_p}{\bf F}_{dd}^{ij}$, while the interactions
with obstacles are given by
${\bf F}_{\rm obs}^i=\sum_j^{N_{ps}}{\bf F}_{dd}^{ij}$.
We apply a uniform driving force ${\bf F}_d=F_d{\bf \hat x}$
to all mobile disks, with $F_d=0.5$.
We initialize the system by placing $N_p+N_{ps}$ disks of reduced
radius in randomly chosen nonoverlapping positions, and then gradually
expanding the radii to size $R_d$
while allowing all disks to move.
This produces a randomized packing
of homogeneous density  with no internal tensions.
We then randomly assign $N_{ps}$ of the disks to be obstacles, and apply an
external driving force.
After a fixed simulation time of $1 \times 10^6$ simulation time steps,
we determine whether the system has reached a clogged or jammed state based on
whether the average disk velocity $V$ has dropped to zero.

\end{document}